\documentclass[final,5p,times,twocolumn]{elsarticle}

\usepackage{graphicx}
\usepackage{amsfonts}
\usepackage{amsmath}
\usepackage{float}
\usepackage{amssymb,amsbsy}
\usepackage{epstopdf}
\usepackage{xcolor}
\usepackage[normalem]{ulem}
\usepackage{braket}
\usepackage{combelow}
\usepackage{hyperref}

\def\bk{{\mathbf{k}}}

\def\fk{f_\bk}

\allowdisplaybreaks

\journal{Physics Letters B}

\begin{document}

\begin{frontmatter}

\title{Shakhov-type extension of the relaxation time approximation \\
in relativistic kinetic theory and second-order fluid dynamics}

\author[goethe,uvt]{Victor E. Ambru\cb{s}}
\ead{victor.ambrus@e-uvt.ro}
\author[goethe,uvt,wroc]{Etele Moln\'ar}
\address[goethe]{Institut f\"ur Theoretische Physik,
Johann Wolfgang Goethe--Universit\"at,
Max-von-Laue-Str.\ 1, D--60438 Frankfurt am Main, Germany}
\address[uvt]{Department of Physics, West University of Timi\cb{s}oara,
Bd.~Vasile P\^arvan 4, Timi\cb{s}oara 300223, Romania}
\address[wroc]{Incubator of Scientiﬁc Excellence--Centre for Simulations of Superdense Fluids,\\
University of Wroc{\l}aw, pl. M. Borna 9, PL-50204 Wroc{\l}aw, Poland
}

\begin{abstract}
We present a relativistic Shakhov-type generalization of the Anderson-Witting
relaxation time model for the Boltzmann collision integral to modify the ratio of momentum
diffusivity to thermal diffusivity.
This is achieved by modifying the path on which the single particle distribution function $f_{\bk}$ approaches local equilibrium $f_{0\bk}$ by constructing an intermediate Shakhov-type distribution $f_{{\rm S} \bk}$ similar to the 14-moment approximation of Israel and Stewart.
We illustrate the effectiveness of this model in case of the Bjorken expansion of an ideal gas of massive particles and the damping of longitudinal waves through an ultrarelativistic ideal gas.
\end{abstract}

\end{frontmatter}

\date{\today}

\section{Introduction}\label{sec:intro}

The relativistic Boltzmann equation, $k^{\mu} \partial_\mu f_\bk = C[f]$,
determines the space-time evolution of a local invariant distribution function
$f_\bk = f(k^\mu,x^\mu)$, where $C[f]$ is the collision term specifying the interaction among constituents.
The relaxation time approximation (RTA), introduced by Anderson and Witting (AW) \cite{Anderson:1974a,Anderson:1974b}
as a proper relativistic generalization of the Bhatnagar-Gross-Krook (BGK) model \cite{Bhatnagar.1954}, approximates the main features of $C[f]$ for both analytic and numeric computations \cite{Florkowski:2013lya,Florkowski:2014sfa,Denicol:2014xca,Bazow.2016,Denicol:2018pak,McNelis:2021zji,Kurkela:2015qoa,Heller:2018qvh,Kurkela:2020wwb,Schlichting:2019abc,Berges:2020fwq,Ambrus:2017keg,Ambrus:2022adp}.
In the AW model the approach to local equilibrium is given by a momentum independent relaxation time, $\tau_R$, while the conservation of a given quantum charge, and of energy and momentum is strictly valid choosing the local rest frame and matching conditions of Landau~\cite{Landau.2014}.
Such relaxation-time models share the common caveat that all first-order transport coefficients are related to $\tau_R$.
In the BGK model, the Prandtl number $\mathrm{Pr}$, representing a ratio of viscosity to thermal diffusivity, is fixed at $1$, while most ideal gases have $\mathrm{Pr} = 2/3$.
This limitation was remedied in the extension proposed by Shakhov~\cite{shakhov68a,shakhov68b} by introducing another parameter that allows $\mathrm{Pr}$ to be controlled independently of $\tau_R$.
This simple modification already leads to a remarkably good agreement with the solutions of the Boltzmann equation and with experimental data, while the BGK model shows significant deviations from both \cite{sharipov1998data,sharipov2002application,graur2009comparison,Li:2015}.

In this paper, we introduce the relativistic Shakhov model as an extension of
the AW model, offering additional energy-independent relaxation times that are related to the corresponding
first-order transport coefficients, the bulk viscosity $\zeta$, particle diffusion $\kappa$ and shear viscosity $\eta$.
This modified collision term, significantly extends the applicability of kinetic RTA models
and allows to study different parameterizations of the first-order transport coefficients
in various physical phenomena such as the baryon diffusion~\cite{Denicol:2018wdp},
and the study of bulk and shear viscosities~\cite{Gardim:2020mmy,Hirvonen:2022xfv} in high-energy heavy-ion collisions.

The improvements compared to the AW model will be shown by comparing the numerical solutions of the kinetic model with solutions of  Israel-Stewart-type second-order fluid dynamics in two different benchmarks: the (0+1)--dimensional longitudinally boost-invariant flow of massive particles highlighting the $\zeta/\eta$ ratio, and the damping of sound waves for an ultrarelativistic fluid concerning the $\kappa/\eta$ ratio.

We note that an energy-dependent relaxation time related to the microscopic details of the collisional processes~\cite{Dusling:2009df,Dusling:2011fd,Kurkela:2017xis} will also modify the first-order transport coefficients~\cite{Rocha:2021zcw,Rocha.2021,Rocha:2022fqz,Dash:2021ibx,Dash:2023ppc}.
For example, setting $\tau_R(x^\mu, E_\bk) = \tau_R(x^\mu) E_\bk^\ell$ in the AW model provides an extra parameter $\ell$ that can be used to tune either $\zeta / \eta$ or $\kappa / \eta$, while it is not excluded that more elaborate models could tune both of these ratios simultaneously.
The Shakhov model introduced in this paper provides an alternative to such AW models with energy-dependent relaxation times.

The detailed derivation of second-order fluid dynamics and the entropy production from the Shakhov model as well as the numerical methods are relegated to the supplementary material.

\section{The Anderson-Witting model and transport coefficients}
\label{sec:AW}

The Boltzmann equation in the Anderson-Witting relaxation-time approximation \cite{Anderson:1974a,Anderson:1974b} for
the binary collision integral reads
\begin{equation}
 k^\mu \partial_\mu f_\bk = C_{\rm AW}[f], \quad
 C_{\rm AW}[f] \equiv
 -\frac{E_\bk}{\tau_R} \delta f_\bk,
 \label{eq:AW}
\end{equation}
where $E_\bk = k^\mu u_\mu $ is the comoving energy of a particle with four-momentum $k^\mu$ and rest mass $m_0^2 = k^\mu k_\mu$, while $\tau_R \equiv \tau_R(x^\mu)$ is the local momentum-independent relaxation time.
The fluid four-velocity $u^{\mu}=\gamma (1,\mathbf{v})$ is normalized to $u^{\mu}u_{\mu}= 1$, where $\gamma=(1-\mathbf{v}^{2})^{-1/2}$.
Here, $\delta f_\bk \equiv f_\bk - f_{0\bk}$ denotes the deviation of $f_\bk$ from its local equilibrium form, the J\"uttner distribution~\cite{Juttner},
\begin{equation}
f_{0\mathbf{k}}= \left[ \exp \left( \beta k^\mu u_\mu - \alpha\right) + a\right]^{-1},
\end{equation}
where $\alpha=\mu\beta$ and $\beta=1/T$ is the inverse temperature, $\mu$ is the chemical potential, while $a=\pm 1$ for fermions/bosons and $a = 0$ for classical Boltzmann particles.

The fluid four-velocity defining the local rest frame is chosen as $u^\mu = T^{\mu\nu} u_\nu/(u_\alpha T^{\alpha \beta} u_\beta)$, where $T^{\mu\nu} = \int dK k^\mu k^\nu f_\bk$ is the energy-momentum tensor, with $dK = [g / (2\pi)^3] d^3k / k^0$ denoting the Lorentz-invariant integration measure in momentum space and $g$ is the degeneracy factor.
With respect to $u^\mu$, $T^{\mu\nu}$ and the particle four-flow $N^\mu = \int dK k^\mu f_\bk$ are decomposed as
\begin{align}
 N^\mu = N_0^\mu + V^\mu, \quad
 T^{\mu\nu} = T^{\mu\nu}_0 - \Pi\Delta^{\mu\nu} + \pi^{\mu\nu},
\end{align}
where $\Delta^{\mu\nu} = g^{\mu\nu} - u^\mu u^\nu$ is the projector orthogonal to $u^\mu$ and
$g^{\mu \nu}=\text{diag}(+,-,-,-)$ is the metric tensor. The bulk viscous pressure $\Pi$, the particle diffusion current $V^\mu$, and the shear-stress tensor $\pi^{\mu\nu}$ represent dissipative corrections due to $\delta f_\bk$.
$N_0^\mu$ and $T^{\mu\nu}_0 $ correspond to the local equilibrium state defined as
\begin{equation}
 N^\mu_0 \equiv \int dK k^\mu f_{0\bk} = nu^\mu, \quad
 T^{\mu\nu}_0 \equiv \int dK k^\mu k^\nu  f_{0\bk} = e u^\mu u^\nu - P \Delta^{\mu\nu}.
\end{equation}
Here, $n \equiv N^{\mu}u_{\mu} = N^{\mu}_0 u_{\mu}$ is the particle density,
$e \equiv T^{\mu \nu} u_\mu u_\nu = T^{\mu \nu}_0 u_\mu u_\nu$ is the energy density, while $P\equiv -T^{\mu\nu}_0\Delta_{\mu \nu}/3 = P(e,n)$ is the thermodynamic pressure defined by an equation of state.

To evaluate the first-order transport coefficients $\zeta$, $\kappa$ and $\eta$, we apply the Chapman-Enskog method \cite{Cercignani.2002}, hence from Eq.~\eqref{eq:AW}, $\delta f_\bk$ is considered to be of first order with respect to $\tau_R$,
\begin{equation}
 \delta f_\bk \simeq -\frac{\tau_R}{E_\bk} k^\mu \partial_\mu f_{0\bk}.
 \label{eq:AW_CE}
\end{equation}
The irreducible moments of $\delta f_\bk$ are defined as~\cite{Denicol:2012cn}:
\begin{equation}
	\rho_{r}^{\mu _{1}\cdots \mu _{\ell}} \equiv \int dK E_\bk^r
	k^{\langle \mu_1} \cdots k^{\mu_\ell \rangle}  \delta f_\bk ,
	\label{rho_r_general}
\end{equation}
where $r$ denotes the power of energy $E_{\mathbf{k}}$ and
$k^{\left\langle \mu _{1}\right. }\cdots k^{\left. \mu _{\ell}\right\rangle }
=\Delta_{\nu_{1}\cdots \nu_{\ell }}^{\mu_{1}\cdots \mu_{\ell }}k^{\nu_{1}}\cdots k^{\nu_{\ell }}$
are the irreducible tensors forming an orthogonal basis \cite{Denicol:2012cn,Groot.1980}.
The symmetric and traceless projection tensors of rank $2\ell$, $\Delta_{\nu_{1}\cdots \nu_{\ell }}^{\mu_{1}\cdots\mu_{\ell }}$, are orthogonal to $u^\mu$ since they are constructed using
the $\Delta^{\mu \nu }$ projector operators.

Through the irreducible moments, $\Pi \equiv -m_0^2\rho_{0}/3$,
$V^{\mu} \equiv \rho_{0}^{\mu}$, and $\pi^{\mu\nu} \equiv \rho_{0}^{\mu \nu}$, we obtain the corresponding first-order approximations for the dissipative quantities,
\begin{gather}
 \Pi = -\zeta_{\rm R} \theta, \quad
 V^\mu = \kappa_{\rm R} \nabla^\mu \alpha, \quad
 \pi^{\mu\nu} = 2\eta_{\rm R}\sigma^{\mu\nu},
 \label{eq:AW_constitutive}
\end{gather}
where $\nabla^\mu = \Delta^\mu_\nu \partial^\nu$ is the gradient operator, $\theta = \nabla^{\mu} u_\mu$
is the expansion rate, and $\sigma^{\mu\nu} = \nabla^{\langle \mu}u^{\nu \rangle}$ is the shear stress tensor.
To arrive at the above results, one splits the space-time derivative $\partial_\mu = u_\mu D + \nabla_\mu$ in Eq.~\eqref{eq:AW_CE} into the comoving derivative $D = u^\mu \partial_\mu$ and the spatial gradient $\nabla_\mu$.
Furthermore, the comoving derivatives of $n$, $u^\mu$ and $T$ are replaced from the conservation equations
of ideal fluid dynamics~\cite{Cercignani.2002}.
The transport coefficients are proportional to $\tau_R$:
\begin{equation}
	\zeta_{\rm R} = \frac{m_0^2}{3} \tau_R \alpha^{(0)}_0, \quad
	\kappa_{\rm R} = \tau_R \alpha^{(1)}_0 , \quad \eta_{\rm R} = \tau_R \alpha^{(2)}_0 ,
	\label{eq:AW_tcoeffs}
\end{equation}
where, for an ideal gas with conserved particle number, the $\alpha^{(\ell)}$ coefficients are
\begin{gather}\label{eq:alphas}
	\alpha^{(0)}_r = -\beta J_{r+1,1} + \frac{1}{G_{22}}[(e + P) G_{2r} - n G_{3r}], \nonumber\\
	\alpha^{(1)}_r = J_{r+1,1} - \frac{n}{(e+P)} J_{r+2,1}, \quad
	\alpha^{(2)}_r = \beta J_{r+3,2}.
\end{gather}
When the particle number is not conserved, i.e., $\mu \equiv 0$, then~\cite{Ambrus:2023qcl}
\begin{equation}
	\alpha^{(0)}_r = -\beta J_{r+1,1} + \frac{J_{31}}{J_{30}} \beta J_{r+1,0}.
	\label{eq:alpha0r_mu0}
\end{equation}
Furthermore $G_{nm} = J_{n0} J_{m0} - J_{n-1,0} J_{m+1,0}$ and the thermodynamic integrals with $\tilde{f}_{0\bk} = 1-af_{0\mathbf{k}}$ are defined as:
\begin{equation}
	J_{nq} = \frac{(-1)^q}{(2q+1)!!} \int dK\, E_\bk^{n-2q} \left(\Delta^{\alpha\beta} k_\alpha k_\beta \right)^q f_{0\bk} \tilde{f}_{0\bk}. \label{eq:Jnq}
\end{equation}

\section{The relativistic Shakhov-type model}\label{sec:Shk}

The main idea behind the Shakhov-type model \cite{shakhov68a,shakhov68b} is to replace the collision
term~\eqref{eq:AW} with
\begin{equation}
 C_{\rm S}[f] \equiv -\frac{E_\bk}{\tau_R}(\fk - f_{{\rm S}\bk}) = -\frac{E_\bk}{\tau_R}(\delta f_\bk - \delta f_{{\rm S}\bk}),
 \label{eq:shk}
\end{equation}
where $f_{{\rm S}\bk} = f_{0\bk} + \delta f_{{\rm S}\bk}$
replaces $f_{0\bk}$ in the AW model.
This collision term drives $f_\bk$ towards $f_{{\rm S}\bk}$, on a time-scale given by $\tau_R$,
and ultimately towards $f_{0\bk}$ on a modified path compared to $ C_{\rm AW}[f]$ from Eq.~\eqref{eq:AW}.
We will construct $f_{{\rm S}\bk}$ such that $\delta f_{{\rm S}\bk}$ vanishes equivalently
in local equilibrium when $\delta f_\bk = 0$.
As in the AW model, we also consider that the relaxation time is independent of the four-momentum of particles. Note that an energy-dependent generalization of the RTA was studied in Refs.~\cite{Rocha:2021zcw,Rocha.2021,Rocha:2022fqz,Dash:2021ibx,Dash:2023ppc}.

Applying the Chapman-Enskog method while considering both $\delta f_{{\rm S}\bk}$ and $\delta f_\bk$ of first-order with respect to $\tau_R$, Eq.~\eqref{eq:AW_CE} is modified to
\begin{equation}
 \delta f_\bk - \delta f_{{\rm S}\bk} \simeq -\frac{\tau_R}{E_\bk} k^\mu \partial_\mu f_{0\bk}.
 \label{eq:Shk_CE}
\end{equation}
Alike to Eq.~\eqref{rho_r_general}, we define the irreducible moments of $\delta f_{{\rm S}\bk}$,
\begin{equation}
	\rho^{\mu_1 \cdots \mu_\ell}_{{\rm S},r} \equiv \int dK E_\bk^r
	k^{\langle \mu_1} \cdots k^{\mu_\ell \rangle} \delta f_{{\rm S}\bk}.
	\label{rhoS_r_general}
\end{equation}
The first-order relations corresponding to Eq.~\eqref{eq:Shk_CE} now read
\begin{gather}
 \Pi - \Pi_{\rm S} = -\zeta_{\rm R} \theta, \quad
 V^\mu - V_{\rm S}^\mu = \kappa_{\rm R} \nabla^\mu \alpha, \nonumber\\
 \pi^{\mu\nu} - \pi_{\rm S}^{\mu\nu} = 2 \eta_{\rm R} \sigma^{\mu\nu}.
 \label{eq:shk_constitutive_aux}
\end{gather}
Note that $\Pi_{\rm S} \equiv -m^2_0 \rho_{S,0}/3$,
$V^{\mu}_{\rm S} \equiv \rho_{S,0}^{\mu}$, and $\pi^{\mu\nu}_{\rm S} \equiv \rho_{S,0}^{\mu \nu}$
are so far unspecified. We aim to use these new irreducible moments to modify the transport coefficients to the following form:
\begin{equation}
	\zeta = \zeta_{\rm R} \frac{\tau_\Pi}{\tau_R},  \quad
	\kappa =  \kappa_{\rm R}  \frac{\tau_V}{\tau_R}, \quad
	\eta = \eta_{\rm R} \frac{\tau_\pi }{\tau_R} ,
	\label{eq:shk_tcoeffs}
\end{equation}
where $\tau_\Pi$, $\tau_V$ and $\tau_\pi$ are the new $\tau_R$-independent relaxation times of bulk viscosity, particle diffusion and shear viscosity, representing parameters of the Shakhov model.
In other words, we seek to replace Eqs.~\eqref{eq:AW_constitutive} by
\begin{gather}
	\Pi = -\zeta \theta, \quad
	V^\mu = \kappa \nabla^\mu \alpha,  \quad
	\pi^{\mu\nu} = 2 \eta \sigma^{\mu\nu},
	\label{eq:shk_constitutive}
\end{gather}
and hence we solve Eqs.~\eqref{eq:shk_constitutive_aux} and \eqref{eq:shk_constitutive} by setting~\cite{shakhov68a,shakhov68b}
\begin{align}
 \Pi_{\rm S} &= \Pi\left(1 - \frac{\tau_R}{\tau_\Pi}\right), &
 V_{\rm S}^\mu &= V^\mu\left(1 - \frac{\tau_R}{\tau_V}\right), &
 \pi_{\rm S}^{\mu\nu} &= \pi^{\mu\nu}\left(1 - \frac{\tau_R}{\tau_\pi}\right).
 \label{eq:dissipative_new}
\end{align}

The conservation equations $\partial_\mu N^\mu =0$ and $\partial_\mu T^{\mu\nu} = 0$ are
fulfilled when
\begin{equation}
 u_\mu N^\mu = u_\mu N^\mu_{\rm S} , \quad
 u_\mu T^{\mu\nu} = u_\mu T^{\mu\nu}_{\rm S},
 \label{eq:Shk_cons}
\end{equation}
where $N_{\rm S}^\mu = \int dK k^\mu f_{{\rm S}\bk}$ and
$T^{\mu\nu}_{\rm S} = \int dK k^\mu k^\nu  f_{{\rm S}\bk}$.
Imposing Landau's matching conditions $n = n_{\rm S}$ and $e = e_{\rm S}$
defines the local chemical potential and temperature. The local rest frame and $u^\mu$ is chosen according to Landau's definition and leads to $\Delta^\mu_\nu u_\lambda T^{\nu \lambda}_{\rm S} = 0$.
Note however that in the Shakhov model other definitions for $u^\mu$ are possible, as pointed out using a similar construction in Ref.~\cite{Pennisi.2018}.

The conditions from Eqs.~\eqref{eq:dissipative_new} and \eqref{eq:Shk_cons} can be written in terms of the irreducible moments
\begin{align}
 \rho_{{\rm S},0} &\equiv -\frac{3}{m_0^2} \Pi_{\rm S}, &
 \rho_{{\rm S},1} &\equiv n_{\rm S} - n = 0, &
 \rho_{{\rm  S},2} &\equiv e_{\rm S} - e = 0,\nonumber\\
 \rho_{{\rm S},0}^\mu &\equiv V^\mu_{\rm S}, &
 \rho_{{\rm S},1}^\mu &\equiv 0, &
 \rho_{{\rm S},0}^{\mu\nu} &\equiv \pi^{\mu\nu}_{\rm S},
 \label{eq:Shk_rhos}
\end{align}
while all other irreducible moments of $\delta f_{{\rm S}\bk}$ are unconstrained.

In order to satisfy Eqs.~\eqref{eq:Shk_rhos},
we construct the relativistic Shakhov distribution $f_{{\rm S}\bk}$ similarly to the near-equilibrium expansion of Grad \cite{Grad:1949} and Israel-Stewart \cite{Israel.1979}.
Starting from
\begin{equation}
f_{{\rm S}\bk} \equiv f_{0\bk} + \delta f_{{\rm S}\bk}
= f_{0\bk} + f_{0\bk} \tilde{f}_{0\bk} \mathbb{S}_\bk,
\end{equation}
we define the Shakhov term $\mathbb{S}_{\bk}$ as:
\begin{align} \nonumber
	\mathbb{S}_\bk &\equiv -\frac{3}{m_0^2}\Pi_{\rm S}
	\mathcal{H}^{(0)}_{\bk 0}
	+ k_{\langle \mu \rangle} V^\mu_{\rm S} \mathcal{H}^{(1)}_{\bk 0}
	+ k_{\langle\mu} k_{\nu \rangle} \pi^{\mu\nu}_{\rm S} \mathcal{H}^{(2)}_{\bk 0} \\ \nonumber
	&= -\frac{3}{m_0^2} \Pi\left(1 - \frac{\tau_R}{\tau_\Pi}\right)
	\mathcal{H}^{(0)}_{\bk 0}
	+ k_{\langle \mu \rangle} V^\mu \left(1 - \frac{\tau_R}{\tau_V}\right) \mathcal{H}^{(1)}_{\bk 0} \\
	&+ k_{\langle\mu} k_{\nu \rangle} \pi^{\mu\nu} \left(1 - \frac{\tau_R}{\tau_\pi}\right)
	\mathcal{H}^{(2)}_{\bk 0}.
	\label{eq:shk_gen}
\end{align}
The polynomials $\mathcal{H}^{(\ell)}_{\bk 0}$ in energy $E_\bk$ are constructed to fulfill Eqs.~\eqref{eq:Shk_rhos} and were introduced in Ref.~\cite{Denicol:2012cn}.
In the $14$--moment approximation, these lead to:
\begin{align}
\mathcal{H}_{\bk 0}^{(0)}=& \frac{G_{33} - G_{23} E_\bk + G_{22} E_\bk^2}{J_{00} G_{33} - J_{10} G_{23} + J_{20} G_{22}}, \nonumber\\
\mathcal{H}_{\bk 0}^{(1)} =& \frac{J_{31} E_\bk - J_{41}}{J_{21} J_{41} - J_{31}^2}, \quad
\mathcal{H}_{\bk 0}^{(2)} = \frac{1}{2 J_{42}}.
\label{eq:Hfunctions}
\end{align}%

\section{Second-order fluid dynamics} \label{sec:hydro}

The Chapman-Enskog method defines the dissipative
fields relating the thermodynamic forces through first-order transport coefficients,
resulting in the Navier-Stokes relations, Eqs.~\eqref{eq:AW_constitutive} and~\eqref{eq:shk_constitutive}.
The method of moments by Grad \cite{Grad:1949}, and Israel and Stewart \cite{Israel.1979}, based on an expansion around equilibrium,
alike the Shakhov distribution, leads to relaxation-type equations
of motion for the dissipative fields $\Pi$, $V^{\mu}$ and $\pi^{\mu \nu}$.

Using the well known $14$--moment approximation of dissipative fluid dynamics
\cite{Denicol:2012cn,Israel.1979,Denicol:2012es,Ambrus:2022vif},
the second-order relaxation equations provide closure for the conservation laws.
Such relaxation equations can be derived for the Shakhov model,
where the additional microscopic time scales introduced through $\delta f_{{\rm S}\bk}$
correspond to the relaxation times $\tau_\Pi$, $\tau_V$, and $\tau_\pi$
from a linearized collision integral, e.g. in the case of binary hard-sphere interactions \cite{Denicol:2012cn,Wagner:2023joq}.
Consequently, some second-order transport coefficients acquire different values than
in the case of the AW model, as shown in the Supplementary Material \cite{SM}.

The equations of motion for the irreducible moments $\rho^{\mu_1 \cdots \mu_\ell}_r$ are derived from the Boltzmann equation for an arbitrary collision term. Here, we summarize them up to rank 2, see Eqs.~(35)--(37) of Ref.~\cite{Denicol:2012cn}:
\begin{subequations}\label{rhodot_all}
	\begin{align}
		\dot{\rho}_r - C_{{\rm S},r-1} &= \alpha^{(0)}_r \theta + \text{higher-order terms}, \label{rhodot}\\
		\dot{\rho}^{\langle\mu\rangle}_r - C^\mu_{{\rm S},r-1} &= \alpha^{(1)}_r \nabla^\mu \alpha
		+ \text{higher-order terms}, \label{rhodot_mu}\\
		\dot{\rho}^{\langle\mu\nu\rangle}_r - C^{\mu\nu}_{{\rm S},r-1} &= 2\alpha^{(2)}_r \sigma^{\mu\nu}
		+ \text{higher-order terms}.\label{rhodot_mu_nu}
	\end{align}
\end{subequations}
The irreducible moments of the Shakhov collision term $C_{\rm S}[f]$ from Eq.~\eqref{eq:shk} are
\begin{equation}
	C^{\mu_1 \cdots \mu_\ell}_{{\rm S},r-1} \equiv \int dK C_{\rm S}[f] E_\bk^{r-1} k^{\langle \mu_1} \cdots k^{\mu_\ell\rangle} = -\frac{1}{\tau_R} \rho^{\mu_1 \cdots \mu_\ell}_r + \frac{1}{\tau_R} \rho^{\mu_1 \cdots \mu_\ell}_{{\rm S},r},
	\label{Cr}
\end{equation}
where the first term on the right hand side corresponds to the AW collision term, while
the second term involves the irreducible moments of $\delta f_{{\rm S}\bk}$.

The collision matrix $\mathcal{A}^{(\ell)}_{rn}$ is defined by~\cite{Denicol:2012cn,Wagner:2023joq}
\begin{equation}
	C^{\mu_1 \cdots \mu_\ell}_{{\rm S},r-1} = -\sum_n \mathcal{A}^{(\ell)}_{rn} \rho^{\mu_1 \cdots \mu_\ell}_n,
\end{equation}
where the summation over $n$ is unrestricted from below or from above. Using Eq.~\eqref{Cr}, the collision matrix evaluates to
\begin{equation}
	\mathcal{A}^{(\ell)}_{rn} = \frac{\delta_{rn}}{\tau_R} - \frac{\delta_{n0}}{\tau_R} \left(1 - \frac{\tau_R}{\tau_{\rm S}^{(\ell)}}\right) \mathcal{F}^{(\ell)}_{-r,0} ,
	\label{Arn}
\end{equation}
where according to Eq.~(66) of Ref.~\cite{Denicol:2012cn}
\begin{equation}
	\mathcal{F}^{(\ell)}_{rn} = \frac{\ell!}{(2\ell + 1)!!} \int dK\, f_{0\bk} \tilde{f}_{0\bk} E_\bk^{-r} (\Delta^{\alpha\beta} k_\alpha k_\beta)^\ell \mathcal{H}^{(\ell)}_{\bk n},
	\label{eq:Frn_def}
\end{equation}
is such that $\mathcal{F}^{(\ell)}_{-r,n} = \delta_{rn}$ for all $0 \le r,n \le N_\ell$, with $\mathcal{H}^{(\ell)}_{\bk n}$ given in Eqs.~\eqref{eq:Hfunctions} and $(N_0, N_1, N_2) = (2,1,0)$.
The different relaxation times are denoted by $\tau_{\rm S}^{(\ell)}$,
with $\tau_{\rm S}^{(0)} = \tau_\Pi$, $\tau^{(1)}_{\rm S} = \tau_V$ and $\tau^{(2)}_{\rm S} = \tau_\pi$.
The inverse of the collision matrix is the relaxation-time matrix, satisfying $\sum_m \tau^{(\ell)}_{rm} \mathcal{A}^{(\ell)}_{mn} = \delta_{rn}$, and it is given by
\begin{equation}
	\tau^{(\ell)}_{rn} = \delta_{rn} \tau_R - \delta_{n0} \left(\tau_R - \tau^{(\ell)}_{\rm S} \right) \mathcal{F}^{(\ell)}_{-r, 0},
	\label{taurn}
\end{equation}
where the elements corresponding to the matching conditions and the choice of frame, $\rho_1=\rho_2=0$ and $\rho^\mu_1=0$, are excluded.

Multiplying Eqs.~\eqref{rhodot_all} by the inverse collision matrix $\tau^{(\ell)}_{nr}$ and summing over $r$ leads to
\begin{subequations}\label{rhodot_aux_all}
	\begin{align}
		\sum_{r \neq 1,2} \tau^{(0)}_{nr} \dot{\rho}_r + \rho_n
		&= \frac{3}{m_0^2} \zeta_n \theta + \text{higher-order terms}, \label{SM:rhodot_aux}\\
		\sum_{r\neq 1} \tau^{(1)}_{nr} \dot{\rho}^{\langle\mu\rangle}_r + \rho_n^{\mu}
		&= \kappa_n \nabla^\mu \alpha + \text{higher-order terms}, \label{SM:rhodot_aux_mu}\\
		\sum_r \tau^{(2)}_{nr}  \dot{\rho}^{\langle\mu\nu\rangle}_r + \rho_n^{\mu\nu}
		&= 2\eta_n \sigma^{\mu\nu} + \text{higher-order terms}, \label{SM:rhodot_aux_mu_nu}
	\end{align}
\end{subequations}
where the first-order transport coefficients are defined as
\begin{align}
	\zeta_r &= \frac{m_0^2}{3} \sum_{n \neq 1, 2} \tau^{(0)}_{rn} \alpha^{(0)}_n, &
	\kappa_r &= \sum_{n \neq 1} \tau^{(1)}_{rn} \alpha^{(1)}_n, &
	\eta_r &= \sum_n \tau^{(2)}_{rn} \alpha^{(2)}_n.
	\label{eq:tcoeffs_def}
\end{align}
Now using Eq.~\eqref{taurn} in the above formulas, we get
\begin{subequations}\label{tcoeffs}
	\begin{align}
		\zeta_r &= \frac{m_0^2}{3} \left[\tau_R \alpha^{(0)}_r - \left(\tau_R - \tau_\Pi \right) \mathcal{F}^{(0)}_{-r,0} \alpha^{(0)}_0\right],\label{SM:zetar}\\
		\kappa_r &= \tau_R \alpha^{(1)}_r - \left(\tau_R - \tau_V \right) \mathcal{F}^{(1)}_{-r,0} \alpha^{(1)}_0, \label{SM:kappar}\\
		\eta_r &= \tau_R \alpha^{(2)}_r - \left(\tau_R - \tau_\pi \right) \mathcal{F}^{(2)}_{-r,0} \alpha^{(2)}_0,  \label{SM:etar}
	\end{align}
\end{subequations}
thus the first-order transport coefficients corresponding to $r = 0$ reduce to
\begin{equation}
	\zeta = \frac{m_0^2}{3} \tau_\Pi \alpha^{(0)}_0, \quad
	\kappa = \tau_V \alpha^{(1)}_0, \quad
	\eta = \tau_\pi \alpha^{(2)}_0,
\end{equation}
in agreement with Eqs.~\eqref{eq:AW_tcoeffs} and \eqref{eq:shk_tcoeffs}.
Contrasting these results with the transport coefficients of the
AW model, the Shakhov model allows a more general approach analogous to the 14-moment approximation of second-order fluid dynamics.
Note that, aside from the first-order transport coefficients in Eqs.~\eqref{eq:shk_tcoeffs}, the Shakhov term in Eq.~\eqref{eq:shk_gen} also affects some second-order transport coefficients.
These are discussed in the Supplementary Material \cite{SM}.

We will illustrate the capabilities of the relativistic Shakhov model by comparing it to the Anderson-Witting model and to second-order fluid dynamics in two different settings.

\begin{figure*}[t]
\begin{center}
\begin{tabular}{ccc}
 \includegraphics[width=.6\columnwidth]{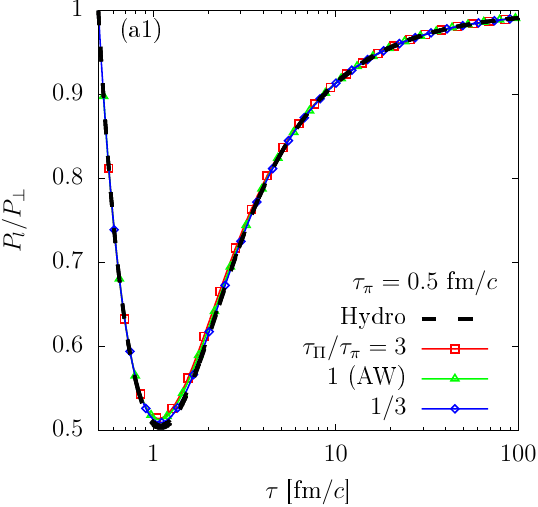} &
 \includegraphics[width=.6\columnwidth]{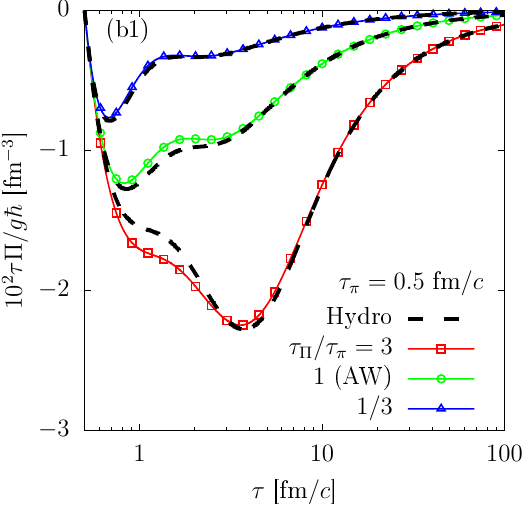} &
 \includegraphics[width=.6\columnwidth]{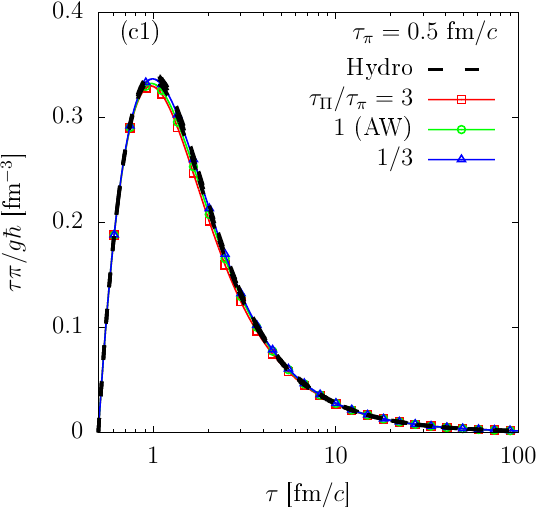} \\
 \includegraphics[width=.6\columnwidth]{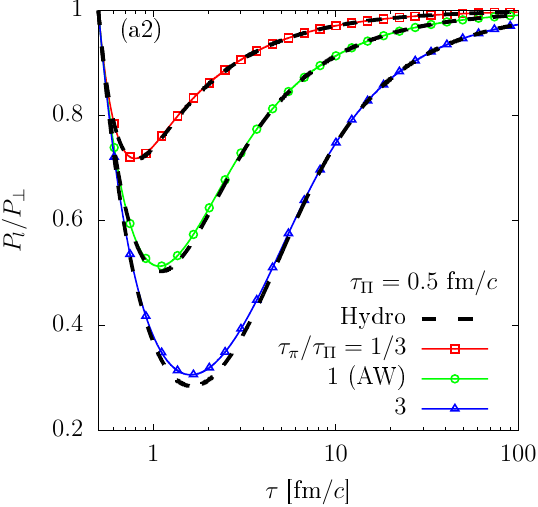} &
 \includegraphics[width=.6\columnwidth]{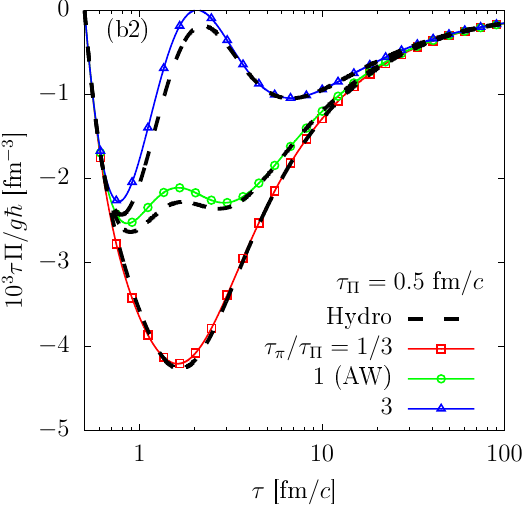} &
 \includegraphics[width=.6\columnwidth]{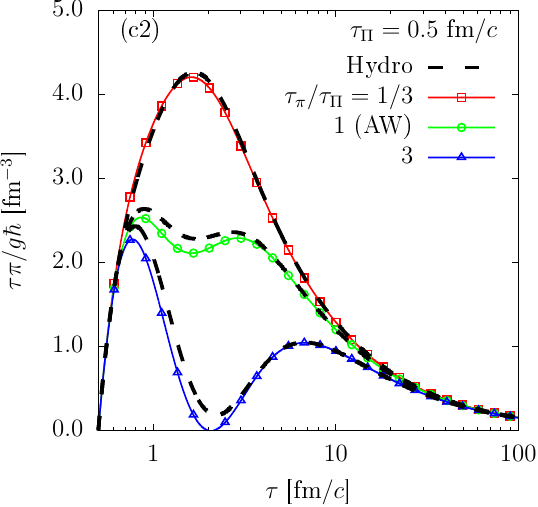} \\
\end{tabular}
\end{center}
\caption{
Time evolution of (a1,a2) $P_l/P_\perp$, (b1,b2), $\tau \Pi / g \hbar$, and (c1,c2) $\tau \pi / g \hbar$ in a $0+1$-dimensional Bjorken
expansion of an ideal gas with vanishing chemical potential and rest mass $m_0 = 1\ {\rm GeV}/c^2$.
In panels (a1)--(c1), $\tau_\pi = 0.5\ {\rm fm} / c$ is fixed and $\tau_\Pi / \tau_\pi$ is changed via the Shakhov collision term. In panels (a2)--(c2), $\tau_\Pi =0.5 \ {\rm fm} / c$ and $\tau_\pi / \tau_\Pi$ is varied.
The value $\tau_\Pi / \tau_\pi = 1$ corresponds to the AW model (green line), while $\tau_\Pi / \tau_\pi = 3$ and $1/3$ are shown with red and blue lines, respectively.
The dashed black lines are the solutions of second-order fluid dynamics in
Eqs.~(\ref{eq:bjork_edot},\ref{eq:bjork_hydro}).
\label{fig:bjork}
}
\end{figure*}

\section{Example I: Bjorken flow} \label{sec:bjork}

We consider the $(0+1)$--dimensional boost-invariant Bjorken expansion~\cite{Bjorken:1982qr} of a classical ($a \equiv 0$)
ideal gas of massive particles. Enforcing $\mu \equiv \nabla^\mu \alpha = 0$ leads to $f_{0\bk} = \exp(-\beta E_\bk)$ and $\partial_\mu N^\mu \neq 0$, with vanishing diffusion current.
This leaves only two independent transport coefficients: $\zeta$ and $\eta$.
Now, replacing $\tau_R=\tau_\Pi$ and  $J_{42} = (e + P)/\beta^2$, the Shakhov distribution~\eqref{eq:shk_gen} reduces to
\begin{equation}
	f_{{\rm S} \bk} =
	 f_{0\bk} \left[1 + \frac{\beta^2 k_{\langle \mu} k_{\nu \rangle} \pi^{\mu\nu}}{2(e + P)} \left(1 - \frac{\tau_\Pi}{\tau_\pi}\right)\right].
	\label{eq:shk_bjork}
\end{equation}
In the $(\tau, x, y, \eta_s)$ coordinates, with proper-time $\tau =\sqrt{t^2 - z^2}$ and space-time rapidity $\eta_s= {\rm artanh}(z/t)$, the energy-momentum tensor is diagonal, $T^{\mu\nu} = {\rm diag}(e, P_\perp, P_\perp, \tau^{-2} P_l)$. The transverse $P_\perp$ and longitudinal $P_l$ pressure components are related to the pressure $P$, the bulk pressure $\Pi$, and the shear-stress tensor component $\pi = \pi^{\eta_s}_{\eta_s}$ via
\begin{equation}
 P_\perp = P + \Pi + \frac{\pi}{2},\quad
 P_l = P + \Pi - \pi.
\end{equation}
In the Shakhov model, the time evolution of $P$, $\Pi$ and $\pi$ is obtained by directly solving the Boltzmann equation $k^\mu \partial_\mu f = C_{\rm S}[f]$, as described in the Supplementary Material \cite{SM}.

For direct comparison to the kinetic model, we also compute the fluid-dynamical evolution, where the energy conservation equation $u_\nu \partial_\mu T^{\mu\nu} = 0$ reduces to
\begin{equation}
 \tau \frac{d e}{d\tau} + e + P_l = 0,
 \label{eq:bjork_edot}
\end{equation}
and it is closed by the following relaxation equations~\cite{Ambrus:2023qcl,Denicol:2014mca}
\begin{subequations}\label{eq:bjork_hydro}
\begin{align}
	\tau \frac{d \Pi}{d\tau} + \left(\frac{\delta_{\Pi\Pi}}{\tau_\Pi} + \frac{\tau}{\tau_\Pi}\right) \Pi - \frac{\lambda_{\Pi\pi}}{\tau_\Pi} \pi &= -\frac{\zeta}{\tau_\Pi},\label{eq:bjork_hydro_bulk}\\
	\tau \frac{d \pi}{d\tau} + \left(\frac{\delta_{\pi\pi}}{\tau_\pi} + \frac{\tau_{\pi\pi}}{3\tau_\pi} + \frac{\tau}{\tau_\pi}\right) \pi - \frac{2\lambda_{\pi\Pi}}{3\tau_\pi} \Pi &= \frac{4\eta}{3\tau_\pi},
	\label{eq:bjork_hydro_shear}
\end{align}
\end{subequations}
with the first-order transport coefficients given by
\begin{equation}
 \zeta = \tau_\Pi \frac{m_0^2 \beta}{3} \left(\frac{J_{31}}{J_{30}} J_{10} -J_{11}\right), \quad
 \eta = \tau_\pi \beta J_{32}. \label{eq:bjork_tcoeffs}
\end{equation}
The second-order transport coefficients $\delta_{\Pi\Pi}$, $\lambda_{\Pi\pi}$, $\delta_{\pi\pi}$, $\tau_{\pi\pi}$, and $\lambda_{\pi\Pi}$ are derived from the Shakhov collision term, as shown in the Supplementary Material \cite{SM}.

In order to validate the kinetic Shakhov model, we performed two sets of simulations with particles of mass $m_0 = 1\ {\rm GeV} / c^2$. At initial time $\tau_0 = 0.5\ {\rm fm} / c$ and initial temperature $\beta_0^{-1} = 0.5\ {\rm GeV}$, we set $f_{\bk}(\tau_0) = \exp(-\beta_0 E_\bk)$. The results obtained using the AW model with $\tau_\pi = \tau_\Pi = \tau_R = 0.5\ {\rm fm} / c$, leading to $4 \pi \eta / s \simeq 2.6$ at the initial time, are used as reference.
The solutions of the corresponding second-order fluid-dynamical equations \eqref{eq:bjork_edot}--\eqref{eq:bjork_hydro} are shown in Fig.~\ref{fig:bjork} with dashed black lines.

Panels (a1)--(c1) of Fig.~\ref{fig:bjork} show the results when
$\tau_\pi = 0.5\ {\rm fm} / c$ is fixed, while $\tau_\Pi = \tau_R$ is varied.
The red, green and blue lines with symbols correspond to the kinetic equation with $\tau_\Pi /\tau_\pi = 3$, $1$ and $1/3$, respectively.
Panels (a1) and (c1) show that the evolution of $P_l / P_\perp$ and $\pi$ remains independent of $\tau_\Pi$.
Panel (b1) shows that the amplitude of $\Pi$ scales roughly with the ratio $\tau_\Pi / \tau_\pi$.
Panels (a2)--(c2) depict the results for fixed $\tau_\Pi \equiv \tau_R = 0.5\ {\rm fm} / c$ and variable $\tau_\pi$. During an initial period of time, (b2) indicates a non-linear dependence of $\Pi$ on $\tau_\pi$. However, at late times, all curves converge towards an asymptotic solution governed by the value of $\tau_\Pi$.  Panels (a2) and (c2) display the expected dependence of $P_l / P_\perp$ and $\pi$ on $\tau_\pi$, with a larger dip and a slower approach to equilibrium, $P_l / P_\perp = 1$ and $\pi = 0$, for larger values of $\tau_\pi$.
The small discrepancies observable in Fig.~\ref{fig:bjork} between the kinetic and fluid-dynamical results are expected due to the large values of the relaxation times and their associated viscosities \cite{Denicol:2014mca,El:2009vj,Bouras:2009nn}. The results at smaller/larger viscosities are in better/worse agreement.

\begin{figure}[t]
\begin{center}
\begin{tabular}{c}
 \includegraphics[width=.9\columnwidth]{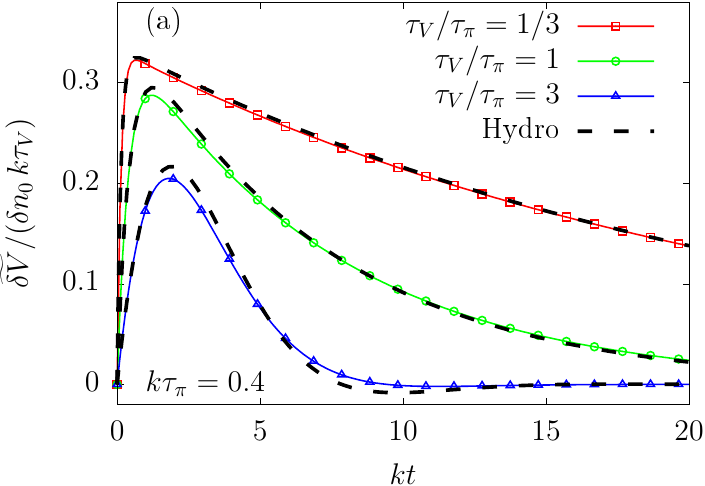}\\
 \includegraphics[width=.9\columnwidth]{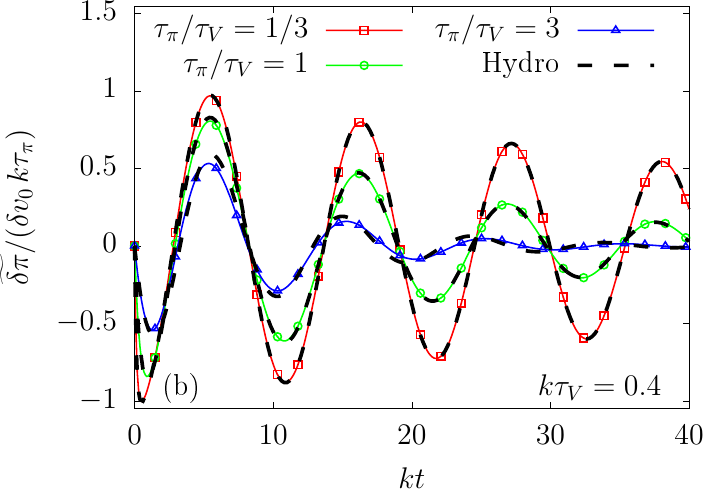}
\end{tabular}
\end{center}
\caption{
Time evolution of the amplitudes (a) $\widetilde{\delta V} / (\delta n_0 k \tau_V)$ for fixed $k\tau_\pi = 0.4$ and variable $\tau_V / \tau_\pi$; and (b) $\widetilde{\delta \pi} / (\delta v_0 k \tau_\pi)$ for fixed $k\tau_V = 0.4$ and variable $\tau_\pi / \tau_V$, in the context of the propagation of longitudinal waves of wave number $k = 2\pi/L$ through an ultrarelativistic gas. The curves with $\tau_V = \tau_\pi$ correspond to the AW model.
The dashed black lines are the solutions of second-order fluid-dynamics. The initial amplitudes were set to $\delta n_0 / n_0 = \delta v_0 = 10^{-3}$.
\label{fig:long}
}
\end{figure}

\section{Example II: Longitudinal waves} \label{sec:long}

We now consider sound waves propagating through an ultrarelativistic, classical ideal gas at rest, with temperature $\beta_0^{-1} = 0.6\ {\rm GeV}$ and $\alpha_0 = 0$.
Since $\zeta = 0$ when $m_0 = 0$, we are left with only two transport coefficients: $\kappa$ and $\eta$. Setting $\tau_\pi = \tau_R$ and $\tau_V \neq \tau_R$ in Eq.~\eqref{eq:shk_gen} and using $J_{nq} = P \beta^{2-n} (n + 1)! / [2(2q+1)!!]$ leads to
\begin{equation}
 f_{{\rm S}\bk} = f_{0\bk} \left[1 + \frac{k_{\langle \mu \rangle} V^\mu}{P} (\beta E_\bk - 5) \left(1 - \frac{\tau_\pi}{\tau_V}\right) \right],
 \label{eq:shk_long}
\end{equation}
where $f_{0\bk} = e^{\alpha - \beta E_\bk}$. At the initial time, the system is in equilibrium and the density $n = n_0 + \delta n$ and velocity $u^z = \delta v$ are perturbed via
$\delta n = \delta n_0 \cos(kz)$ and $\delta v = \delta v_0 \sin(kz)$,
while the pressure is unperturbed, $P = P_0 = n_0 T_0$.
Here, $k = 2\pi / L$ and $L = 6.4\ {\rm fm}$, while $\delta n_0 / n_0 = \delta v_0 / c = 10^{-3}$. We then track the time evolution of the amplitudes $\widetilde{\delta \pi}$ and $\widetilde{\delta V}$, computed as
\begin{equation}
 \widetilde{\delta V} = \frac{2}{L} \int_0^L dz\, \delta V\, \cos(k z), \quad
 \widetilde{\delta \pi} = \frac{2}{L} \int_0^L dz\, \delta \pi\, \sin(k z),
\end{equation}
where $\delta V = V^z$ and $\delta \pi \simeq \pi^{zz}$ correspond to the diffusion
current and shear-stress tensor, respectively. For more details, see Sec.~V of Ref.~\cite{Ambrus:2022vif}.

We now compare the numerical solution of the full Shakhov model described by Eqs.~\eqref{eq:shk} and \eqref{eq:shk_long}, obtained as described in the Supplementary Material \cite{SM}, with the solution of the linearized second-order fluid dynamical equations, computed using the methods of Refs.~\cite{Ambrus:2017keg,Ambrus:2022vif}.
Panel (a) shows the ratio $\widetilde{\delta V} / (\delta n_0 k \tau_V)$ for three values of $\tau_V / \tau_\pi$, with $k\tau_\pi = 0.4$ fixed, such that $4\pi \eta / s \simeq 3.11$. It can be seen that the amplitude of $\widetilde{\delta V}$ and the damping exponent scale almost linearly with $\tau_V$.
Similarly, for the $\widetilde{\delta \pi} / (\delta v_0 k \tau_\pi)$ ratio shown in panel (b),
we varied $\tau_\pi /\tau_V$ while keeping $k \tau_V = 0.4$ fixed. This panel shows that the damping exponent of $\widetilde{\delta \pi}$ is proportional to $\tau_\pi$.
Small discrepancies between the kinetic and fluid-dynamical results when $ \tau_V / \tau_\pi = 3$ in panel (a) or $\tau_\pi / \tau_V = 3$ in panel (b) are due to the departure from the domain of applicability of second-order fluid dynamics \cite{Ambrus:2017keg}.

\section{Conclusion}\label{sec:conc}

In this paper, we introduced the relativistic Shakhov-type model as an extension of the Anderson-Witting RTA with momentum-independent relaxation time.
This model introduces new and independently-adjustable first-order transport coefficients/relaxation times for the bulk viscous pressure, diffusion current, and shear-stress tensor.
Although here our primary focus was on the first-order transport coefficients, the Shakhov model can be systematically extended \cite{Ambrus:2024qsa} to control all transport coefficients of second-order fluid dynamics.

This more general kinetic equation makes it possible to study different values or parameterizations of the first-order transport coefficients in various phenomena such as the baryon diffusion or the effect of bulk and shear viscosities similarly as in the state of the art second-order fluid dynamical simulations.
Therefore, the present work significantly extends the applicability of kinetic RTA models
to describe systems relevant in high-energy heavy-ion collisions.

\vspace{-10pt}
\section*{Acknowledgements}
\vspace{-10pt}
We thank P.~Huovinen, H.~Niemi, D.~Wagner, S.~Busuioc, A.~Dash, P.~Aasha and D.~H.~Rischke for the
reading of the manuscript.
This work was supported through a grant of the
Ministry of Research, Innovation and Digitization, CNCS - UEFISCDI,
project number PN-III-P1-1.1-TE-2021-1707, within PNCDI III.
V.E.A.~acknowledges the support of the Alexander von Humboldt
Foundation through a Research Fellowship for postdoctoral researchers.
E.M.~was supported by the program Excellence Initiative--Research
University of the University
of Wroc{\l}aw of the Ministry of Education
and Science.

{\it Code availability.--}
The numerical code, raw data, and scripts to generate the plots shown in this paper are available on Code Ocean~\cite{codeoceanshk}.
The code for the Bjorken flow extends that of Ref.~\cite{Ambrus:2023qcl}, which is also
available at Ref.~\cite{codeoceanbjork}.
The modified Bessel functions $K_n(z)$ and the Bickley function ${\rm Ki}_1(z)$ are computed using the algorithms developed by D.E.~Amos~\cite{AmosBessel,Amos609}, which are available through the OpenSpecfun project. \footnote{Source files downloaded from\\ \texttt{https://github.com/JuliaMath/openspecfun}, commit number\\ \texttt{70239b8d1fe351042ad3321e33ae97923967f7b9}.}
\vspace{-10pt}

\bibliographystyle{unsrtnat}
\bibliography{bib_Shakhov}

\newpage

\begin{center}
 {\huge SUPPLEMENTARY \\[0.3cm] MATERIAL}
\end{center}

\setcounter{equation}{0}
\setcounter{figure}{0}
\setcounter{section}{0}

\renewcommand{\theequation}{SM-\arabic{equation}}
\renewcommand{\thefigure}{SM-\arabic{figure}}
\renewcommand{\thesection}{SM-\arabic{section}}

This supplementary material is structured in three sections. Section~\ref{SM:sec:tcoeffs} discusses the second-order transport coefficients from the Shakhov model. Section~\ref{SM:sec:entropy} presents the entropy production, while Section~\ref{SM:sec:num} summarizes the details of the numerical scheme used to solve the Shakhov model equation.

\section{Second-order transport coefficients of the relativistic Shakhov model}\label{SM:sec:tcoeffs}

In this section we employ the method of moments of
Refs.~\cite{Denicol:2012cn,Ambrus:2023qcl} to derive the first- and second-order transport
coefficients corresponding to the relativistic Shakhov model.
These transport coefficients arise at first- and second-order with respect to the Knudsen number $\textrm{Kn}$, being the ratio of the particle mean free path and a characteristic macroscopic scale, and the inverse Reynolds number $\mathrm{Re}^{-1}$, being the ratio of an out-of-equilibrium and a local-equilibrium macroscopic field.

{\it Irreducible moments and orthogonal basis.--}
The irreducible moments from Eq.~\eqref{rho_r_general} are expressed as~\cite{Denicol:2012cn},
\begin{equation}
\delta f_{\mathbf{k}} = f_{0\mathbf{k}} \tilde{f}_{0\bk}
\sum_{\ell =0}^{\infty }\sum_{n=0}^{N_{\ell }} \rho_{n}^{\mu_{1}\cdots \mu_{\ell }}
k_{\left\langle \mu _{1}\right. }\cdots k_{\left.\mu _{\ell }\right\rangle}
\mathcal{H}_{\mathbf{k}n}^{(\ell)},
\label{SM:f_k_expansion}
\end{equation}
where $N_\ell \rightarrow \infty$ is an expansion order.
The functions $\mathcal{H}^{(\ell)}_{\bk n}$ are polynomials of order $N_\ell$ with respect to $E_\bk$, defined in full generality in Eq. (29) of Ref.~\cite{Denicol:2012cn}, and are constructed such that Eq.~\eqref{rho_r_general} is satisfied for $0 \le r \le N_\ell$.
We remark that, while Eq.~\eqref{SM:f_k_expansion} employs an irreducible basis, the expansion does not account explicitly for the negative-order moments $\rho_r^{\mu_1 \cdots \mu_\ell}$ with $r < 0$, but
these must be reconstructed from those with $0 \le r \le N_\ell$ in a manner which becomes exact only in the limit $N_\ell \rightarrow \infty$.
The simple structure of the RTA model allows us to circumvent such construction in Eq.~\eqref{SM:f_k_expansion} by employing a basis-free approach, as discussed in Ref.~\cite{Ambrus:2022vif}.

We note that the functions $\mathcal{H}^{(\ell)}_{\bk n}$, related to the representation of $\delta f_\bk$  are also useful in the context of the Shakhov model. However, for the Shakhov distribution, $N_\ell$ is not the expansion order of $\delta f_\bk$, but the order of the
$\mathcal{H}^{(\ell)}_{\bk 0}$ polynomials satisfying the constraints in Eq.~\eqref{eq:Shk_rhos},
namely $N_0 = 2$, $N_1 = 1$, and $N_2 = 0$.

The Shakhov collision term from Eq.~\eqref{eq:shk} is
\begin{equation}
 C^{\mu_1 \cdots \mu_\ell}_{r-1} = -\frac{1}{\tau_R} \rho^{\mu_1 \cdots \mu_\ell}_r + \frac{1}{\tau_R} \rho^{\mu_1 \cdots \mu_\ell}_{{\rm S},r},
 \label{SM:Cr}
\end{equation}
where the second term involves the irreducible moments of $\delta f_{{\rm S}\bk} = f_{0\bk} \tilde{f}_{0\bk} \mathbb{S}_\bk$ defined in Eq.~\eqref{rhoS_r_general}.
Now, using the Shakhov distribution from Eq.~\eqref{eq:shk_gen}, leads to
\begin{gather}
 \rho_{{\rm S},r} = -\frac{3\Pi}{m_0^2} \left(1 - \frac{\tau_R}{\tau_\Pi}\right) \mathcal{F}^{(0)}_{-r,0}, \quad
 \rho_{{\rm S},r}^\mu = V^\mu \left(1 - \frac{\tau_R}{\tau_V}\right) \mathcal{F}^{(1)}_{-r,0}, \nonumber\\
\rho_{{\rm S},r}^{\mu\nu} = \pi^{\mu\nu} \left(1 - \frac{\tau_R}{\tau_\pi}\right) \mathcal{F}^{(2)}_{-r,0},\label{SM:rhoS}
\end{gather}
while the higher-rank moments are set to vanish, i.e., $\rho_{{\rm S},r}^{\mu_1 \cdots \mu_\ell}=0$
with $\ell > 2$.
Now, using Eq.~\eqref{eq:Frn_def} for polynomial orders $N_0 = 2$, $N_1 = 1$ and $N_2 = 0$ ensures that
$\mathcal{F}^{(0)}_{0,0} = \mathcal{F}^{(1)}_{0,0} = \mathcal{F}^{(2)}_{0,0} = 1$ and $\mathcal{F}^{(0)}_{-1,0} = \mathcal{F}^{(0)}_{-2,0} = \mathcal{F}^{(1)}_{-1,0} = 0$.

The second-order transport coefficients also require the knowledge of various other moments
$\rho^{\mu_1 \cdots \mu_\ell}_{r \neq 0}$. Here we recall the first-order approximation to such
irreducible moments in the so-called basis-free approach of Ref.~\cite{Ambrus:2022vif}:
\begin{equation}
	\rho_{r\neq 0} \simeq -\frac{3}{m_0^2} \mathcal{R}^{(0)}_{r0} \Pi , \quad
	\rho^\mu_{r \neq 0} \simeq \mathcal{R}^{(1)}_{r0} V^\mu , \quad
	\rho^{\mu\nu}_{r \neq 0} \simeq \mathcal{R}^{(2)}_{r0} \pi^{\mu\nu} , \label{SM:basis_free}
\end{equation}
where
\begin{equation}
	\mathcal{R}^{(0)}_{r0} = \frac{\zeta_r}{\zeta}\, , \quad
	\mathcal{R}^{(1)}_{r0} = \frac{\kappa_r}{\kappa}\, ,\quad
	\mathcal{R}^{(2)}_{r0} = \frac{\eta_r}{\eta}\, . \label{SM:Clr_def}
\end{equation}

Now, substituting the expressions for the first-order transport coefficients from Eqs.~\eqref{tcoeffs} into Eq.~\eqref{SM:Clr_def} gives
\begin{equation}
 \mathcal{R}^{(\ell)}_{-r,0} =  \frac{\tau_R}{\tau_{\rm S}^{(\ell)}}
 \frac{\alpha^{(\ell)}_{-r}}{\alpha^{(\ell)}_0} + \left(1 - \frac{\tau_R}{\tau_{\rm S}^{(\ell)}}\right) \mathcal{F}^{(0)}_{r0} .
 \label{SM:Clr}
\end{equation}
Using these results, the relaxation times can be computed using Eqs.~(38) of Ref.~\cite{Wagner:2022ayd}:
\begin{align}
 \tau_\Pi &= \sum_{r \neq 1,2} \tau^{(0)}_{0r} \mathcal{R}^{(0)}_{r0}, &
 \tau_V &= \sum_{r \neq 1} \tau^{(1)}_{0r} \mathcal{R}^{(1)}_{r0}, &
 \tau_\pi &= \sum_r \tau^{(2)}_{0r} \mathcal{R}^{(2)}_{r0}.
\end{align}
Recalling the expression for $\tau^{(\ell)}_{nr}$ from Eqs.~\eqref{taurn} together with Eq.~\eqref{SM:Clr}, the above definitions lead to $\tau_\Pi = \tau^{(0)}_{\rm S}$, $\tau_V = \tau^{(1)}_{\rm S}$ and $\tau_\pi = \tau^{(2)}_{\rm S}$,
as expected.

As discussed in Ref.~\cite{Ambrus:2022vif}, the second-order transport coefficients involve only the coefficients $\mathcal{R}^{(\ell)}_{-1,0}$ and $\mathcal{R}^{(\ell)}_{-2,0}$.
These coefficients also require the expressions for $\mathcal{F}^{(\ell)}_{r0}$, computed using the functions $\mathcal{H}^{(\ell)}_{\bk 0}$ in Eq.~\eqref{eq:Hfunctions}, as shown below:
\begin{gather}
 \mathcal{F}^{(0)}_{r0} = \frac{J_{-r,0} G_{33} - J_{1-r,0} G_{23} + J_{2-r,0} G_{22}}{J_{00} G_{33} - J_{10} G_{23} + J_{20} G_{22}}, \nonumber\\
 \mathcal{F}^{(1)}_{r0} = \frac{J_{2-r,1} J_{41} - J_{3-r,1} J_{31}}{J_{21} J_{41} - J_{31}^2},\quad
 \mathcal{F}^{(2)}_{r0} = \frac{J_{4-r,2}}{J_{42}}. \label{SM:Frn}
\end{gather}

{\it Equations of motion.--}
The relaxation equations for $\Pi = -m^2_0  \rho_{0}/3$,
$V^{\mu} = \rho_{0}^{\mu}$, and $\pi^{\mu\nu} =\rho_{0}^{\mu \nu}$ are obtained by setting $n=0$ in Eqs.~\eqref{rhodot_aux_all}.
Up to second order with respect to ${\rm Kn}$ and ${\rm Re}^{-1}$, these equations read, see Eqs.~(88-93)
in Ref.~\cite{Ambrus:2022vif},
\begin{subequations}
	\begin{align}
		& \tau_\Pi \dot{\Pi} + \Pi = -\zeta \theta - \ell_{\Pi V} \nabla_\mu V^\mu - \tau_{\Pi V} V_\mu \dot{u}^\mu - \delta_{\Pi \Pi} \Pi \theta \nonumber\\
		& \hspace{30pt} - \lambda_{\Pi V} V_\mu \nabla^\mu \alpha + \lambda_{\Pi \pi} \pi^{\mu\nu} \sigma_{\mu\nu}, \label{SM:Pidot}\\
		& \tau_V \dot{V}^{\langle \mu \rangle} + V^\mu = \kappa \nabla^\mu \alpha - \tau_V V_\nu \omega^{\nu \mu} - \delta_{VV} V^\mu \theta \nonumber\\
		& \hspace{30pt}  - \ell_{V\Pi} \nabla^\mu \Pi + \ell_{V\pi} \Delta^{\mu\nu} \nabla_\lambda \pi^\lambda_\nu + \tau_{V\Pi} \Pi \dot{u}^\mu - \tau_{V\pi} \pi^{\mu\nu} \dot{u}_\nu \nonumber\\
		& \hspace{30pt}   - \lambda_{VV} V_\nu \sigma^{\mu\nu} + \lambda_{V\Pi} \Pi \nabla^\mu \alpha  - \lambda_{V\pi} \pi^{\mu\nu} \nabla_\nu \alpha,\label{SM:Vdot}\\
		& \tau_\pi \dot{\pi}^{\langle \mu\nu \rangle} + \pi^{\mu\nu} = 2 \eta \sigma^{\mu\nu} + 2 \tau_\pi \pi_\lambda^{\langle \mu} \omega^{\nu \rangle \lambda} - \delta_{\pi\pi} \pi^{\mu\nu} \theta \nonumber\\
		& \hspace{30pt}  -\tau_{\pi\pi} \pi^{\lambda \langle\mu} \sigma^{\nu\rangle}_\lambda + \lambda_{\pi \Pi} \Pi \sigma^{\mu\nu} - \tau_{\pi V} V^{\langle \nu} \dot{u}^{\mu \rangle} \nonumber\\
		& \hspace{30pt}  + \ell_{\pi V} \nabla^{\langle \mu} V^{\nu \rangle} + \lambda_{\pi V} V^{\langle \mu} \nabla^{\nu\rangle} \alpha. \label{SM:pidot}
	\end{align}
\end{subequations}

{\it Shakhov model for the Bjorken flow.--}
In the case of the Bjorken expansion, we considered a massive, ideal, uncharged gas, such that $\alpha^{(0)}_r$ is given by Eq.~\eqref{eq:alpha0r_mu0}. The first-order transport coefficients $\zeta$ and $\eta$ are listed in Eqs.~\eqref{eq:bjork_tcoeffs}.
The second-order transport coefficients appearing in Eq.~\eqref{eq:bjork_hydro} are listed here from Ref.~\cite{Ambrus:2023qcl}:
\begin{align}
 \delta_{\Pi\Pi} &= \tau_\Pi \left(\frac{2}{3} + \frac{m_0^2}{3} \frac{J_{10}}{J_{30}}
 + \frac{m_0^2}{3} {\mathcal{R}}^{(0)}_{-2,0}\right), \\
 \lambda_{\Pi \pi} &=  \tau_\Pi\frac{m_0^2}{3} \left(\frac{J_{10}}{J_{30}}
 + {\mathcal{R}}^{(2)}_{-2,0}\right), \quad
 \delta_{\pi\pi} = \tau_\pi \left( \frac{4}{3} + \frac{m_0^2}{3} \mathcal{R}^{(2)}_{-2,0} \right),\nonumber\\
 \tau_{\pi\pi} &= \tau_\pi \left(\frac{10}{7} + \frac{4m_0^2}{7} \mathcal{R}^{(2)}_{-2,0} \right),\quad
 \lambda_{\pi \Pi} = \tau_\pi\left(\frac{6}{5} + \frac{2m_0^2}{5} \mathcal{R}^{(0)}_{-2,0} \right).\nonumber
\end{align}

Since the Shakhov distribution employed in Eq.~\eqref{eq:shk_bjork} uses $\tau_\Pi = \tau_R$, the coefficients $\mathcal{R}^{(0)}_{-r,0}$ reduce to their corresponding values for the AW model, namely
\begin{equation}
	\mathcal{R}^{(0)}_{-r,0} \equiv \frac{\alpha^{(0)}_{-r}}{\alpha^{(0)}_0} =
	\frac{J_{1-r,0} J_{31} - J_{1-r,1} J_{30}}{J_{10} J_{31} - J_{11} J_{30}},
\end{equation}
where Eq.~\eqref{eq:alpha0r_mu0} was employed to replace $\alpha^{(0)}_r$. On the other hand, $\mathcal{R}^{(2)}_{-r,0}$ becomes
\begin{equation}
	\mathcal{R}^{(2)}_{-r,0} = \frac{\tau_\Pi}{\tau_\pi} \frac{J_{3-r,2}}{J_{32}} + \left(1 - \frac{\tau_\Pi}{\tau_\pi}\right) \frac{J_{4-r,2}}{J_{42}},
\end{equation}
which, in the limit of $\tau_\Pi = \tau_\pi$, recovers the analogous coefficient appearing in the AW model, $\alpha^{(2)}_{-r} / \alpha^{(2)}_0 = J_{3-r,2} / J_{32}$.
Therefore, the transport coefficients $\lambda_{\Pi \pi}$, $\delta_{\pi\pi}$, and $\tau_{\pi\pi}$  involving $\mathcal{R}^{(2)}_{-2,0}$ are modified with respect to their AW expressions, while $\delta_{\Pi\Pi}$ and $\lambda_{\pi\Pi}$ remain unchanged.

{\it Shakhov model for longitudinal waves.--}
In the case of the longitudinal waves concerning an ultrarelativistic classical ideal gas, we have
\begin{align}
	J_{nq} &= \frac{P \beta^{2-n} (n+1)!}{2(2q+1)!!}, &
	P &= \frac{g e^\alpha}{\pi^2 \beta^4},\nonumber\\
	\alpha^{(1)}_r &= \frac{P(r+2)!(1-r)}{24 \beta^{r-1}}, &
	\alpha^{(2)}_r &= \frac{P(r+4)!}{30\beta^r}.
\end{align}
The transport coefficients from Eq.~\eqref{SM:Vdot} reduce to~\cite{Ambrus:2022vif}:
\begin{gather}
		\delta_{VV} = \tau_V, \quad
		\ell_{V \pi} = \frac{\tau_V}{h} \!\left( 1- h\mathcal{R}^{(2)}_{-1,0}\right), \quad
		\tau_{V \pi} = \frac{\tau_V}{h} \!
		\left(1 - h\frac{\partial \mathcal{R}^{(2)}_{-1,0}}{\partial \ln \beta}\right), \nonumber\\
		\lambda_{VV} = \frac{3}{5}\tau_V, \quad
		\lambda_{V \pi} = \tau_V \left(\frac{\partial \mathcal{R}^{(2)}_{-1,0}}{\partial \alpha}
		+ \frac{1}{h} \frac{\partial \mathcal{R}^{(2)}_{-1,0}}{\partial \beta} \right),
		\label{SM:diffusion}
\end{gather}
where $h=(e+P)/n$ is the enthalpy per particle.
Noting that
\begin{equation}
 \mathcal{F}^{(2)}_{10} = \frac{\beta}{5}, \quad
 \mathcal{R}^{(2)}_{-1,0} = \frac{\beta}{4}\left(1 + \frac{\tau_R - \tau_\pi}{5\tau_\pi}\right),
\end{equation}
the Shakhov model alters only the following coefficients:
\begin{subequations}\label{SM:UR_vector}
	\begin{align}
		\ell_{V\pi} \equiv \tau_{V\pi} &= \frac{\beta}{20}\left(1 - \frac{\tau_R}{\tau_\pi}\right) \tau_V,\\
		\lambda_{V\pi} &= \frac{\beta}{16}
		\left(1 + \frac{\tau_R - \tau_\pi}{5\tau_\pi}\right) \tau_V.
	\end{align}
Similarly, the coefficients appearing in Eq.~\eqref{SM:pidot} are
	\begin{align}
		\delta_{\pi\pi} &= \frac{4}{3} \tau_\pi, &
		\tau_{\pi\pi} &= \frac{10}{7} \tau_\pi, &
		\ell_{\pi V} &= \tau_{\pi V} = \lambda_{\pi V} = 0, \label{SM:UR_tensor}
	\end{align}
\end{subequations}
while $\kappa = \frac{\beta P}{12} \tau_V$ and $\eta = \frac{4P}{5} \tau_\pi$.
The Shakhov collision term considered in Eq.~\eqref{eq:shk_long} employs $\tau_\pi = \tau_R$, hence the dependence on $\tau_\pi$ disappears in Eqs.~\eqref{SM:UR_vector} and all transport coefficients reduce to the AW ones (with $\tau_R$ replaced by $\tau_V$ or $\tau_\pi$, as appropriate), see for comparison Eqs.~(168) and (169) in Ref.~\cite{Ambrus:2022vif}.

\section{Entropy production} \label{SM:sec:entropy}

We now discuss the thermodynamic consistency of the Shakhov model by considering the entropy production
\begin{equation}
 \partial_\mu S^\mu = -\int dK\, C_{\rm S}[f]  \ln (f_\bk / \tilde{f}_\bk),
 \label{eq:entropy_dS}
\end{equation}
where $S^\mu = -\int dK\, k^\mu (f_\bk \ln f_\bk + a \tilde{f}_\bk \ln \tilde{f}_\bk)$ is the entropy four-current.
As originally pointed out by Shakhov \cite{shakhov68a}, asserting the sign of $\partial_\mu S^\mu$ for arbitrary distributions $f_\bk$ is difficult, but if the fluid is not far from equilibrium, quadratic terms in $\delta f_\bk$ or $\delta f_{{\rm S}\bk}$ can be neglected and the logarithm in Eq.~\eqref{eq:entropy_dS} can be approximated as:
\begin{equation}
 \ln \frac{f_\bk}{\tilde{f}_\bk} = \ln \frac{f_{0\bk}(1 + \tilde{f}_{0\bk} \phi_\bk)}{\tilde{f}_{0\bk} (1 - a f_{0\bk} \phi_\bk)}
 \simeq \ln \frac{f_{0\bk}}{\tilde{f}_{0\bk}} + \phi_\bk + O(\phi_\bk^2),
\end{equation}
where $\phi_\bk = \delta f_\bk /(f_{0\bk} \tilde{f}_{0\bk})$. Thus, Eq.~\eqref{eq:entropy_dS} becomes
\begin{multline}
 \partial_\mu S^\mu \simeq \frac{1}{\tau_R} \int dK E_\bk (\delta f_\bk - \delta f_{{\rm S}\bk}) \ln \frac{f_{0\bk}}{\tilde{f}_{0\bk}} \\
 + \frac{1}{\tau_R} \int dK E_\bk \delta f_\bk \left(\phi_\bk - \mathbb{S}_\bk \right) ,
 \label{eq:entropy_dS_small}
\end{multline}
where on the second line, we have used the relation
$(\delta f_\bk - \delta f_{{\rm S}\bk}) \phi_\bk = \delta f_\bk (\phi_\bk - \mathbb{S}_\bk)$ with
$\mathbb{S}_\bk = \delta f_{{\rm S}\bk}/f_{0\bk} \widetilde{f}_{0\bk}$.
Since $\ln(f_{0\bk} / \tilde{f}_{0\bk}) = \alpha - \beta E_\bk$, the first term on the right-hand side of the equation vanishes due to the matching conditions in Eq.~\eqref{eq:Shk_cons}. The second term can be estimated using Eq.~\eqref{eq:Shk_CE}, leading to
\begin{equation}
 \phi_\bk - \mathbb{S}_\bk \simeq -\frac{\tau_R}{E_\bk} k^\mu \partial_\mu (\alpha - \beta E_\bk),
\end{equation}
and with Eq.~\eqref{eq:shk_constitutive} confirms the second law of thermodynamics,
\begin{equation}
 \partial_\mu S^\mu \simeq \frac{\beta}{\zeta} \Pi^2 - \frac{1}{\kappa} V_\mu V^\mu + \frac{\beta}{2\eta} \pi_{\mu\nu} \pi^{\mu\nu} \ge 0.
\end{equation}

\section{Numerical method for the Shakhov model}
\label{SM:sec:num}

To solve the Shakhov kinetic model $k^\mu \partial_\mu f_\bk = C_{\rm S}[f]$, we employ a discrete velocity method inspired by the Relativistic Lattice Boltzmann algorithm of Refs.~\cite{Ambrus:2022adp,Romatschke:2011hm,Ambrus:2016aub,Gabbana:2017uvc,Gabbana:2019ydb}.
We consider the rapidity-based moments of $f_\bk$ introduced in Ref.~\cite{Ambrus:2023qcl}, which eliminates two out of the three dimensions of the momentum space for the particular case of the $(1+1)$-dimensional longitudinal waves~\ref{app:long}, and the $(0+1)$-dimensional boost invariant expansion~\ref{app:bjork}, respectively.

\subsection{Longitudinal wave damping problem}\label{app:long}

In the application of Sec.~\ref{sec:long}, the fluid is homogeneous with respect to the $x$ and $y$ directions. Parameterizing the momentum space using $(m_\perp, \varphi_\perp, v^z)$ as in Ref.~\cite{Ambrus:2023qcl},
\begin{equation}
	\begin{pmatrix}
		k^t \\ k^z
	\end{pmatrix} = \frac{m_\perp}{\sqrt{1 - v_z^2}}
	\begin{pmatrix}
		1 \\ v^z
	\end{pmatrix}, \quad
	\begin{pmatrix}
		k^x \\ k^y
	\end{pmatrix} = \sqrt{m_\perp^2 - m_0^2}
	\begin{pmatrix}
		\cos\varphi_\perp \\ \sin\varphi_\perp
	\end{pmatrix},
	\label{eq:mperp}
\end{equation}
the Boltzmann equation with the Shakhov model for the collision term reduces to
\begin{equation}
	\partial_t f_\bk + v^z \partial_z f_\bk =
	-\frac{u \cdot v}{\tau_R} (f_\bk - f_{{\rm S}\bk}),\label{eq:boltz_long_aux2}
\end{equation}
where $v^\mu = k^\mu / k^t$ and $u \cdot v = \gamma(1 - \beta^z v^z)$, with $\beta^z$ being the fluid three-velocity along the $z$ direction and $\gamma = 1/\sqrt{1 - \beta_z^2}$. Introducing the rapidity-based moments \cite{Ambrus:2023qcl}
\begin{equation}
	F_n = \frac{g}{(2\pi)^3} \int_0^{2\pi} d\varphi_\perp \int_{m_0}^\infty \frac{dm_\perp\, m_\perp^{n+1}}{(1 - v_z^2)^{(n+2)/2}} f_\bk,
	\label{eq:Fn_def}
\end{equation}
Eq.~\eqref{eq:boltz_long_aux2} becomes
\begin{equation}
	\partial_t F_n + v^z \partial_z F_n = -\frac{u \cdot v}{\tau_R} (F_n - F^{\rm S}_n).
	\label{eq:boltz_Fn_long}
\end{equation}

It can be shown \cite{Ambrus:2024qsa} that the macroscopic quantities $N^t$, $N^z$, $T^{tt}$, $T^{tz}$ and $T^{zz}$ can be obtained from $F_1$ and $F_2$ via
\begin{equation}
	\begin{pmatrix}
		N^t \\ N^z
	\end{pmatrix} = \int_{-1}^1 dv^z
	\begin{pmatrix}
		1 \\ v^z
	\end{pmatrix} F_1, \quad
	\begin{pmatrix}
		T^{tt} \\ T^{tz} \\ T^{zz}
	\end{pmatrix} = \int_{-1}^1 dv^z
	\begin{pmatrix}
		1 \\ v^z \\ v_z^2
	\end{pmatrix} F_2.
\end{equation}
For the case of massless particles considered in Sec.~\ref{sec:long}, $T^\mu{}_\mu = 0$, such that $T^{xx} = T^{yy} = (T^{tt} - T^{zz})/2$.
From the above, it is clear that the time evolution of both $N^\mu$ and $T^{\mu\nu}$ is fully determined by the functions $F_1$ and $F_2$. In order to solve Eq.~\eqref{eq:boltz_Fn_long}, the functions $F^{\rm S}_n$ must be obtained by integrating Eq.~\eqref{eq:shk_long}, yielding:
\begin{equation}
	F^{\rm S}_1 = \frac{n}{2(u \cdot v)^3} - \frac{3V(\beta^z - v^z)}{2(u \cdot v)^4} \left(1 - \frac{\tau_\pi}{\tau_V}\right), \quad
	F^{\rm S}_2 = \frac{3P}{2(u\cdot v)^4}.
\end{equation}

The time discretization is performed using equal time steps $\delta t = 10^{-3}\,{\rm fm}/c$ and the time stepping is performed using the third-order total variation diminishing (TVD) Runge-Kutta scheme \cite{Shu:1988,Gottlieb:1998}.  The spatial domain $[-L/2, L/2]$ is discretized using $S = 100$ cells of size $\delta s = L / S$, centred on $z_s = (s - \frac{1}{2}) \delta s -\frac{L}{2}$, $1 \le s \le S$. The spatial derivative $v^z \partial_z F_n$ is approximated using finite differences:
\begin{equation}
	\left(v_z \frac{\partial F_n}{\partial z}\right)_s = \frac{\mathbf{F}_{n;s+1/2} - \mathbf{F}_{n;s-1/2}}{\delta s},
\end{equation}
where $\mathbf{F}_{n;s+1/2}$ represents the flux at the interface between cells $s$ and $s+1$. For definiteness, we compute this flux using the upwind-biased fifth-order weighted essentially non-oscillatory (WENO-5) scheme introduced in Ref.~\cite{Jiang:1996,Rezzolla.2013}. Finally, the $v^z$ momentum space coordinate is discretized via the Gauss-Legendre quadrature with $K = 20$ points, such that $P_K(v^z_j) = 0$, with $1 \le j \le K$ and $P_K(z)$ being the Legendre polynomial of order $K$. Then, integrals with respect to $v^z$ of a function $g(v^z)$ are approximated via
\begin{equation}
	\int_{-1}^1 dv^z\, g(v^z) \simeq \sum_{j = 1}^K g_j, \qquad
	g_j \equiv w_j g(v^z_j),
\end{equation}
with $w_j$ being the Gauss-Legendre quadrature weights \cite{Hildebrand.1987}.

\subsection{Algorithm for the Bjorken expansion}
\label{app:bjork}

The algorithm for the Bjorken expansion is identical to that described in Ref.~\cite{Ambrus:2023qcl}, hence we only recall the main method here.
The spatial rapidity is $\eta_s = {\rm artanh}(z / t)$, and the parametrization of the momentum space is as in Eq.~\eqref{eq:mperp}, with $(k^t, k^z)$ replaced by $(k^\tau, \tau k^{\eta_s})$, where
\begin{equation}
	k^\tau = \frac{1}{\tau}(t k^t - z k^z), \quad
	\tau k^{\eta_s} = \frac{1}{\tau}(t k^z - z k^t).
\end{equation}
Retaining the definition \eqref{eq:Fn_def} of the rapidity-based moments, the non-vanishing components of $T^{\mu\nu} = {\rm diag}(e, P_\perp, P_\perp, \tau^{-2} P_l)$ are given by \cite{Ambrus:2023qcl}
\begin{align}
	\begin{pmatrix}
		e \\ P_l
	\end{pmatrix} &= \int_{-1}^1 dv^z \begin{pmatrix}
		1 \\ v_z^2
	\end{pmatrix} F_2, &
	P_\perp &= \int_{-1}^1 dv^z \left(\frac{1 - v_z^2}{2} F_2 - \frac{m_0^2}{2} F_0\right).
\end{align}

The Boltzmann equation becomes
\begin{equation}
	\partial_\tau F_n + \frac{1 + (n-1) v_z^2}{\tau} F_n - \frac{1}{\tau} \frac{\partial[v^z(1 - v_z^2) F_n]}{\partial v^z} = -\frac{1}{\tau_R} (F_n - F^{\rm S}_n),
\end{equation}
where $F^{\rm S}_n$ can be obtained by integrating Eq.~\eqref{eq:shk_bjork}:
\begin{align}
	F^{\rm S}_n &= F^{\rm eq}_n - \frac{\beta^2 \pi(1 - \frac{\tau_\Pi}{\tau_\pi})}{4(e + P)} \left[m_0^2 F^{\rm eq}_n - (1 - 3v_z^2) F^{\rm eq}_{n+2}\right],
\end{align}
where $F_n^{\rm eq} = \frac{g}{4\pi^2} \Gamma(n+2,\zeta) / \beta^{n+2}$,
$\Gamma(n,\zeta) = \int_\zeta^\infty dx\, x^{n-1} e^{-x}$ is the incomplete Gamma function and $\zeta = \beta m_0 / \sqrt{1 - v_z^2}$. Since $\pi = \frac{2}{3}(P_\perp - P_l)$, the system is closed in terms of $F_0$ and $F_2$.

The time integration and $v^z$ discretization proceed as in the previous subsection, however now the time step $\delta \tau_n = \tau_{n+1} - \tau_n$ is determined adaptively via
\begin{equation}
	\delta \tau_n = {\rm min}(\alpha_\tau \tau_n, \alpha_R\tau_R),
\end{equation}
with $\alpha_\tau = 10^{-3}$ and $\alpha_R= 1/2$. Furthermore, the derivative with respect to $v^z$ is performed by projecting $F_n$ onto the space of Legendre polynomials, as described in Ref.~\cite{Ambrus:2016aub}. Considering the discretization with $K = 20$ discrete velocities $1 \le j \le K$, we have
\begin{equation}
	\left\{ \frac{\partial[v^z(1 - v_z^2) F_n]}{\partial v^z}\right\}_j = \sum_{j' = 1}^K \mathcal{K}_{j,j'} F_{n;j'},
\end{equation}
where the kernel $\mathcal{K}_{j,j'}$ is given in Eq.~(3.54) of Ref.~\cite{Ambrus:2016aub}.

\end{document}